\documentclass[twocolumn,10pt]{article}

\usepackage[T1]{fontenc}
\usepackage{lmodern}
\usepackage{microtype}
\usepackage{amsmath,amssymb}
\usepackage{graphicx}
\usepackage{hyperref}

\usepackage{makecell}
\usepackage{listings}
\usepackage{dirtytalk}
\usepackage{arydshln}
\usepackage[numbers]{natbib}
\usepackage{subcaption}

\begin{document}
%-------------------------------------------------------------------------------

%don't want date printed
\date{}

\title{\Large \bf Breaking Darknet CAPTCHAs with general purpose LLM}

\author{
{\rm Benjamin Fehrensen}\\
Bern University of Applied Science (BFH)
\and
{\rm Jens Hubler}\\
Bern University of Applied Science (BFH)
} 

\maketitle

%-------------------------------------------------------------------------------
\begin{abstract}
%-------------------------------------------------------------------------------
Our work evaluates the effectiveness of automated methods for solving CAPTCHA challenges commonly encountered in darknet environments. These CAPTCHAs are typically designed to operate without JavaScript, resulting in distinct characteristics compared to mainstream CAPTCHA systems. Our study considers three representative challenge types: open-circle localization, rotation-based alignment, and object-selection CAPTCHAs.

We compare classical image processing techniques with off-the-shelf multimodal large language models (MLLMs) and investigate hybrid architectures that combine both approaches. Our results show that classical image registration methods substantially outperform MLLMs on tasks requiring precise geometric reasoning, such as localization and rotation estimation. In particular, open-circle and rotation CAPTCHAs can be solved reliably and several orders of magnitude faster using deterministic image processing algorithms based on gradient-domain analysis and circular cross-correlation. In contrast, MLLMs excel at semantically complex challenges such as object-selection CAPTCHAs, where robust visual recognition is required.

Our experiments further reveal a systematic limitation of contemporary MLLMs: while they are generally capable of identifying relevant visual structures, they frequently struggle with precise spatial localization and geometric transformations. These deficiencies can be mitigated either through task reformulation or by augmenting the models with specialized image processing tools.

These deficiencies can be mitigated by task reformulation or by equipping the model with specialized image-processing tools. We therefore propose a hybrid framework in which an MLLM serves as a high-level reasoning and orchestration layer while delegating geometric computations to deterministic algorithms via the Model Context Protocol (MCP). The resulting system achieves success rates above 90\% across all evaluated CAPTCHA types and demonstrates that combining the complementary strengths of MLLMs and classical computer vision yields a more accurate and efficient solver than either approach alone.

\end{abstract}

%-------------------------------------------------------------------------------
\section{Introduction}
\label{sect:introduction}
%-------------------------------------------------------------------------------

During the investigation of websites hosting illicit or questionable content, we observed that CAPTCHAs~\cite{GossweilerRich2009WuCa,GuerarMeriem2022GC’A} are frequently deployed as a primary access control mechanism. Notably, the user base of such platforms tends to be highly sensitive to tracking techniques, particularly those relying on JavaScript-based fingerprinting. As a consequence, many users deliberately disable JavaScript in their browsers. To maintain accessibility under these constraints, such websites often avoid requiring JavaScript entirely. This design choice results in a distinct subset of CAPTCHA implementations that differ significantly from those commonly encountered on mainstream websites.

The problem of automatic CAPTCHA solving has been extensively studied, with a wide range of approaches proposed in the literature~\cite{KumarMohinder2022ASSo}. For instance, \cite{breaking_reCAPTCHA} demonstrated that advanced machine learning techniques can achieve high success rates in solving CAPTCHAs, including near-perfect performance against Google reCAPTCHA challenges. More recently, \cite{WangHong2023BoHD,deng2024oedipusllmenchancedreasoningcaptcha, DengGelei2025OLRC, 309468, wang2025cognitionevaluationdefensemultimodal} showed that modern multimodal large language models (MLLMs) are capable of solving CAPTCHA tasks effectively, even without task-specific training.

In this work, we deliberately avoid training custom models and instead focus on evaluating existing methods and systems. Additionally, we exclude the use of CAPTCHA-solving services (so-called CAPTCHA farms), as they are not suitable for our application scenario due to ethical, legal, and scalability considerations.

We further restrict our threat model by excluding attacks that rely on compromising the underlying web service or exploiting implementation vulnerabilities in CAPTCHA systems. Similarly, brute-force approaches targeting the CAPTCHA generation mechanism (e.g., exhaustively enumerating all $2^9$ possible solutions in simplified schemes) are not considered within the scope of this study.

Instead, we adopt a hybrid approach that combines classical image processing techniques, such as gradient-domain cross-correlation, with the capabilities of off-the-shelf MLLM to construct an effective CAPTCHA-solving system.

The contributions of this paper are summarized as follows:

\begin{itemize}
    \item We present a systematic evaluation of CAPTCHA types commonly encountered on darknet platforms, including open-circle, rotation-based, and object-selection challenges.

    \item We compare classical image processing techniques and state-of-the-art multimodal large language models (MLLMs) with respect to accuracy, latency, and token consumption, identifying the strengths and limitations of each approach.

    \item We show that contemporary MLLMs are highly effective at semantic image understanding but exhibit systematic weaknesses in geometric reasoning tasks, such as precise localization and mental rotation.

    \item We demonstrate that these limitations can be mitigated either through task reformulation or by augmenting MLLMs with specialized image processing tools.

    \item We propose and evaluate a hybrid CAPTCHA-solving framework based on the Model Context Protocol (MCP), in which an MLLM acts as an orchestration layer while delegating geometric computations to dedicated image processing algorithms.
\end{itemize}

\section{Related Works}

With the advent of MLLMs and vision-language models (VLMs), recent work has begun to treat CAPTCHA solving as general interactive reasoning tasks.

Our work is inspired and extends on approaches using general purpose pre-trained MLLM to solve CAPTCHAs \cite{WangHong2023BoHD, deng2024oedipusllmenchancedreasoningcaptcha, 309468, DengGelei2025OLRC, wang2025cognitionevaluationdefensemultimodal}. 
Our work is very much focused on special group of CAPTCHAs not using JavaScript.
Arguably simpler to solve as they cannot combine multiple channels such as behaviour based interaction next to visual challenges.
\cite{wang2025cognitionevaluationdefensemultimodal} showed that direct prompts i.e. forwarding the original instructions does not work very well for most CAPTCHA types.
Therefore, right from the beginning we tried to use optimised system and user prompt to improve the success rate.

While recent agentic systems such as Halligan\cite{309468} and Oedipus\cite{deng2024oedipusllmenchancedreasoningcaptcha} demonstrate that general-purpose VLMs can solve a wide range of previously unseen CAPTCHA types through planning, tool use, and multi-step reasoning, they still rely primarily on the VLM itself for both high-level interpretation and low-level geometric computation. Our experiments on darknet-style, JavaScript-free CAPTCHAs reveal a complementary limitation: contemporary MLLMs frequently identify the correct semantic structures (open circles, alignment cues, insects) but fail at precise spatial localization and mental rotation.

We therefore adopt a hybrid design in which the MLLM acts strictly as an orchestration and reasoning layer. Geometric subtasks—open-circle detection via Hough transforms and gap analysis, and rotation estimation via radial gradient profiles and circular cross-correlation—are delegated to deterministic classical image-processing algorithms exposed through the Model Context Protocol (MCP). This separation yields perfect accuracy on the geometric challenges at latencies orders of magnitude lower than pure MLLM or agentic approaches, while preserving the MLLM’s strengths on semantic object-selection tasks. Task reformulation (e.g., presenting pre-rotated candidates instead of requiring angle regression) further mitigates geometric weaknesses when pure algorithmic solutions are unavailable.

In short, pure agentic VLM solvers treat geometric reasoning as another capability to be elicited from the model; our hybrid framework treats it as a capability better offloaded to specialized, reliable tools. The two approaches are complementary: agentic systems excel at open-ended, interactive commercial CAPTCHAs, whereas classical+MLLM hybrids are particularly effective for the structurally constrained, geometry-heavy challenges common in no-JavaScript environments.

\section{Methods and Experiments}

To maximize the success rate of automatically solving CAPTCHA challenges, we investigated several complementary approaches and evaluated their effectiveness across different CAPTCHA types.

\begin{itemize}

\item \textbf{Classical image processing techniques.}  
We applied traditional image analysis and registration methods, including gradient-domain cross-correlation, to solve CAPTCHA challenges algorithmically. These approaches proved particularly effective for structurally simple CAPTCHA types such as open-circle and rotation-based challenges, where geometric alignment and pattern matching are sufficient to recover the correct solution efficiently.

\item \textbf{Prompt engineering for MLLMs.}  
We evaluated the capability of multimodal large language models (MLLMs) to solve CAPTCHA challenges using carefully designed prompts. Both system-level and user-level prompts were iteratively refined to improve task comprehension and response consistency. Since the models were accessed programmatically through Python APIs, we additionally constrained the output format using structured schemas defined via \texttt{pydantic} \texttt{BaseModel} classes, enabling reliable parsing of the generated responses.

\item \textbf{Task simplification and reformulation.}  
We explored transformations of CAPTCHA challenges into representations that are easier for MLLMs to interpret. For example, in rotation-based CAPTCHAs, several models were able to identify the relevant visual features but struggled to estimate rotational angle. To simplify the task, instead of requesting a rotation angle prediction, we generated a discrete set of pre-rotated candidate images (e.g., $45^\circ$, $90^\circ$, $135^\circ$, etc.) and asked the model to select the correctly aligned variant. This reformulation substantially improved robustness and accuracy.

\item \textbf{Hybrid MLLM-assisted solving with external tools.}  
Finally, we augmented the MLLMs with access to classical image processing utilities during inference. This approach is particularly beneficial for CAPTCHA types where deterministic image registration techniques outperform purely semantic reasoning. The external tools were integrated using the Model Context Protocol (MCP), enabling the models to invoke specialized image processing operations as part of the solving pipeline.

\end{itemize}

\subsection{Models}
To improve reproducibility, all experiments were conducted with the model \texttt{temperature} parameter fixed to zero, thereby minimizing stochastic variation in the generated outputs. Additionally, the \texttt{seed} parameter was fixed to the value $42$ to further improve determinism across repeated runs. The \texttt{top\_p} parameter was retained at its default value of $1.0$. To ensure consistent experimental conditions, we restricted our evaluation to MLLMs that explicitly support configuration of these inference parameters.

As a consequence of this constraint, our experiments relied on the OpenAI GPT-4o model rather than newer 5.x-series models, which no longer expose direct control over the \texttt{temperature} parameter.

All evaluated MLLMs were tested using identical prompts, preprocessing steps, and inference parameters. We acknowledge that several vendors provide specialized vision-oriented or premium multimodal models that may achieve superior performance on CAPTCHA-related tasks. However, the objective of this work was not to identify the strongest commercially available model, but rather to investigate how general-purpose off-the-shelf MLLMs can be effectively adapted and optimized for CAPTCHA solving through prompt engineering, task reformulation, and hybrid integration with classical image processing methods.

We evaluate four representative MLLMs obtained from three major providers (Table~\ref{tab:models}). To specifically investigate the impact of explicit reasoning capabilities, we include both a reasoning-enabled and a non-reasoning variant of the Grok-4.20 model.

\begin{table}
\caption{MLLMs evaluated in this study. The selected models represent three major providers and include both reasoning-enabled and non-reasoning configurations. Snapshot identifiers correspond to the model versions used during the experiments to improve reproducibility.}
\label{tab:models}
        
\begin{tabular}{lll}
\hline
Provider & Model & Snapshots\\
\hline
OpenAI 		& GPT-4o  				& 2024-08 \\
Google 		& Gemini-3.1-Pro 			& 2026-02\\
SpaceXAI 	& Grok 4.20-non-reasoning	& 2026-04 \\
SpaceXAI 	& Grok 4.20-reasoning		& 2026-04 \\
\hline
\end{tabular}
\end{table}

\subsection{Evaluation Metrics}

We evaluate the CAPTCHA-solving approaches along three primary dimensions:

\begin{itemize}
    \item \textbf{Accuracy}, measured as the percentage of CAPTCHA challenges solved successfully.
    
    \item \textbf{Latency}, measured as the end-to-end response time from submitting a CAPTCHA challenge to receiving a complete and parseable solution. For MLLM-based approaches, this includes network communication, model inference, and response generation. For classical image processing approaches, it corresponds to the total execution time of the solving pipeline.
    
    \item \textbf{Token consumption}, measured as the total number of input and output tokens processed by the MLLM for each CAPTCHA challenge. This metric serves as a proxy for computational cost and API usage.
\end{itemize}

Together, these metrics allow us to assess not only the effectiveness of each approach, but also its practical efficiency and operational cost.

\subsection{Dataset}

\begin{figure*}[t]
\centering
\begin{subfigure}[t]{0.32\textwidth}
  \centering
  \includegraphics[width=\linewidth]{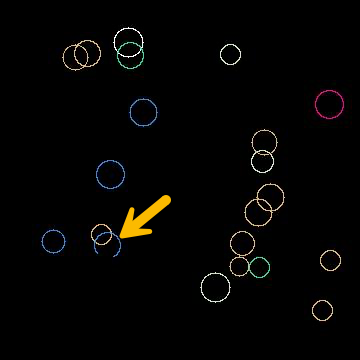}
  \caption{Open-circle CAPTCHA. The yellow arrow points to the open circle.}
  \label{fig:open_circle_sample}
\end{subfigure}\hfill
\begin{subfigure}[t]{0.32\textwidth}
  \centering
  \includegraphics[width=\linewidth]{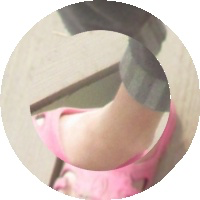}
  \caption{Rotation CAPTCHA. The inner picture has to be rotated by 315° clockwise.}
  \label{fig:rotation_sample}
\end{subfigure}\hfill
\begin{subfigure}[t]{0.32\textwidth}
  \centering
  \includegraphics[width=\linewidth]{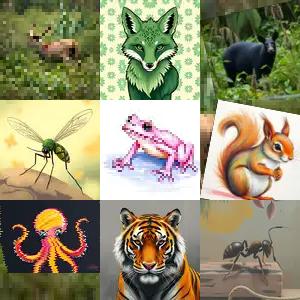}
  \caption{Insects selection CAPTCHA. The squares 4 and 9 contains insects.}
  \label{fig:insects_selection}
\end{subfigure}

\caption{Examples of the three CAPTCHA types considered: open-circle, rotation, and insects selection.}
\label{fig:captcha_examples}
\end{figure*}

We consider three widely used CAPTCHA types commonly encountered in the targeted ecosystem:

\begin{itemize}
    \item \textbf{Open-circle CAPTCHA:} The user is required to click inside an incomplete circular shape (see Figure~\ref{fig:open_circle_sample}).
    \item \textbf{Rotation CAPTCHA:} The task consists of rotating an inner circular image to align it with a surrounding reference image (see Figure~\ref{fig:rotation_sample}).
    \item \textbf{Insects selection CAPTCHA:} The user must identify and select all instances of a specified object class within an image (see Figure~\ref{fig:insects_selection}).
\end{itemize}

All experiments are conducted on a dataset collected from real-world samples obtained from active deployments of these CAPTCHA systems. To ensure reproducibility, the dataset is publicly available at \url{https://github.com/Hu6li/Evaluating_MLLMs_Against_Darknet_CAPTCHAs}.

\section{Results}

We begin by analyzing the results obtained for the open-circle CAPTCHA challenges. These tasks are particularly well suited for classical image processing techniques, where geometric structure and edge information can be exploited effectively.

Next, we examine the rotation-based CAPTCHA challenges. For this category, we hypothesized that both classical image registration methods and MLLMs would perform well. In particular, we expected MLLMs to leverage their strong visual reasoning and object recognition capabilities to infer the correct rotational alignment of the image components.

Finally, we evaluate object-selection CAPTCHAs, such as insect detection tasks. In contrast to the previous CAPTCHA types, these challenges cannot be  addressed using traditional image registration approaches and instead require semantic image understanding. This setting plays directly to the strengths of modern MLLMs, which are known to perform well on visual classification and object recognition tasks.

%%%% OPEN CIRCLE CAPTCHA  =====================================================================================================
\subsection{Open-circle CAPTCHA} 
\label{sec:open-circle}

Several classical image processing approaches are applicable to open-circle CAPTCHAs; however, we describe only the most effective strategy identified during our experiments. The complete implementation is published together with the dataset at \url{https://github.com/Hu6li/Evaluating_MLLMs_Against_Darknet_CAPTCHAs}.

Our approach consists of the following steps:

\begin{enumerate}
    \item \textbf{Image normalization and edge detection.}  
    The input image is first converted to grayscale in order to normalize color variations. Subsequently, edge detection is applied, and all further analysis is performed on the resulting edge representation.

    \item \textbf{Circle candidate detection.}  
    Potential circles are detected using the Hough Circle Transform~\cite{duda1973pattern,bradski2008learning,davies2005machine}, yielding candidate center coordinates and corresponding radii.

    \item \textbf{Gap analysis.}  
    For each detected circle candidate, a radial scan over the full angular range from $0^\circ$ to $360^\circ$ is performed to determine whether the circle contains a gap. Since the estimated center coordinates may deviate slightly from the true center, the scan is evaluated within a tolerance region of $\pm 3$ pixels around the detected radius.

    \item \textbf{False-positive filtering.}  
    Sensitive Hough Transform parameters typically produce multiple spurious circle candidates. Therefore, an additional verification step is applied to determine whether a candidate genuinely represents a circular structure rather than an accidental edge configuration.
\end{enumerate}

\begin{figure*}[t]
\centering
\begin{subfigure}[t]{0.32\textwidth}
  \centering
  \includegraphics[width=\linewidth]{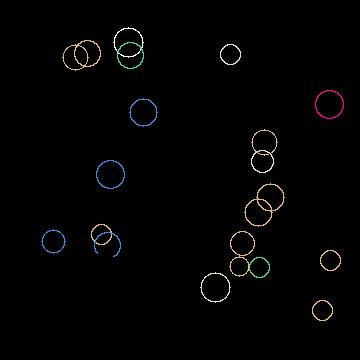}
  \caption{Original open-circle CAPTCHA challenge.}
  \label{fig:org_open_circle}
\end{subfigure}\hfill
\begin{subfigure}[t]{0.32\textwidth}
  \centering
  \includegraphics[width=\linewidth]{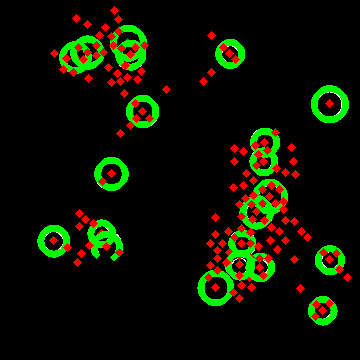}
  \caption{Edge-detected representation used for analysis. Green points denote detected edge pixels on the circle, while red points indicate candidate circle centers identified by the Hough Transform.}
  \label{fig:edges_open_circle}
\end{subfigure}\hfill
\begin{subfigure}[t]{0.32\textwidth}
  \centering
  \includegraphics[width=\linewidth]{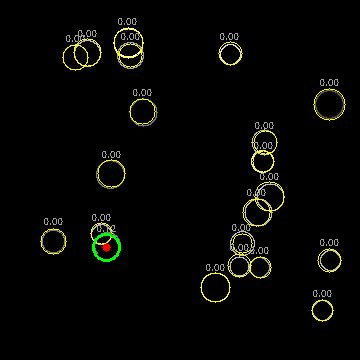}
  \caption{Final detection result. The red point marks the detected center of the open circle, while the green line highlights the detected open circle.}
  \label{fig:detected_open_circle}
\end{subfigure}
\caption{Visualization of the open-circle CAPTCHA solving pipeline using classical image processing techniques and the Hough Circle Transform.}
\label{fig:captcha_examples}
\end{figure*}

Using this comparatively simple image processing pipeline, we already achieved very high success rates. In practice, such CAPTCHA types can therefore be solved reliably without requiring MLLMs at all. Nevertheless, evaluating MLLM performance on these tasks remains interesting, particularly to assess their capability on low-level geometric reasoning problems. Table~\ref{tab:circle-mllm} summarizes the obtained results.

The primary difficulty for MLLMs in solving open-circle CAPTCHAs was not detecting the relevant circle itself, but rather reporting its precise spatial coordinates in a consistent format.

To compensate for potential scaling and resolution differences across images, we instructed the models to return normalized coordinates in the user prompt:

\begin{quote}
\emph{
x\_frac=0.0 corresponds to the left image boundary, while x\_frac=1.0 corresponds to the right boundary. \\
y\_frac=0.0 corresponds to the top image boundary, while y\_frac=1.0 corresponds to the bottom boundary.
}
\end{quote}

Among the evaluated models, Gemini~3.1~Pro was the only MLLM that consistently adhered to this coordinate specification with sufficient precision. To better understand the decision-making process of the models, we additionally required all MLLMs to provide textual reasoning together with their predicted coordinates.

\begin{figure}
  \centering
  \includegraphics[width=0.8\linewidth]{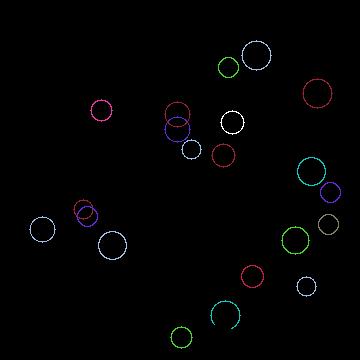}
\caption{Example of a localization failure by Grok-4.20 non-reasoning. The model correctly reported that \emph{``all rings in the image are closed except for one cyan-colored ring located in the lower central area''}. Although the target object was identified correctly, the predicted coordinates were not sufficiently precise to successfully solve the CAPTCHA.}
\label{fig:cyan_ring}
\end{figure}

Inspection of these reasoning traces revealed that the models frequently identified the correct open circle semantically, but failed to accurately localize it spatially. A representative example is shown in Figure~\ref{fig:cyan_ring}. In this case, the Grok-4.20 non-reasoning model correctly recognized that the cyan ring in the lower central region was the only open circle, yet it was unable to return sufficiently accurate coordinates for successful interaction.

Although Gemini~3.1~Pro solved the open-circle CAPTCHAs with near-perfect accuracy, the computational cost and latency are disproportionate to the complexity of the task. Our dedicated image processing pipeline achieves comparable accuracy while being approximately 60 times faster.

The performance advantage of classical image processing techniques naturally raises the question of whether both approaches can be combined. Rather than requiring the MLLM to solve every challenge independently, it can be augmented with specialized tools for tasks that are known to admit efficient algorithmic solutions.

To evaluate this hypothesis, we implemented a hybrid architecture in which the MLLM is provided with access to a \texttt{solve\_open\_circle} tool via MCP. Given a CAPTCHA image, the tool executes the image processing pipeline described above and returns the coordinates of the detected open-circle center. The MLLM can then incorporate this information into its reasoning process and generate the final response without having to perform the underlying geometric analysis itself.

The corresponding results are reported as \emph{Hybrid} in Table~\ref{tab:circle-mllm}. For this task, no measurable difference was observed between the reasoning and non-reasoning variants of Grok-4.20. In all evaluated test cases, both models successfully invoked the MCP tool and returned the coordinates produced by the image processing pipeline without modification. Consequently, the hybrid configurations achieved the same perfect accuracy and localization precision as the standalone image processing approach.

The most notable effect of the hybrid architecture is the substantial improvement in accuracy for models that otherwise performed poorly on this task. GPT-4o improved from 20\% to 100\% accuracy, while both Grok-4.20 variants improved from 10\% to 100\%. This confirms that the primary limitation of these models was not the ability to identify the open circle, but rather the ability to determine and report its precise coordinates.

The hybrid architecture incurs additional token consumption because the model must process both the tool invocation and the information returned by the MCP tool. Nevertheless, increased token usage does not necessarily translate into higher latency. For example, the hybrid Gemini~3.1~Pro configurations consumed substantially more tokens than their standalone counterparts, yet reduced the average response time from 43.9\,s to 8.6\,s and from 88.1\,s to 18.2\,s, respectively. Delegating the geometric localization task to a specialized algorithm therefore proved significantly more efficient than requiring the MLLM to perform the analysis internally.

These results illustrate the complementary strengths of MLLMs and classical image processing techniques. While MLLMs excel at interpreting tasks, selecting appropriate tools, and coordinating workflows, classical image processing algorithms remain substantially more efficient and reliable for precise geometric computations. The hybrid architecture combines these advantages by allowing the MLLM to delegate specialized subtasks to dedicated algorithms whenever appropriate. For the open-circle CAPTCHA, this combination yields both perfect accuracy and substantially improved execution times.

\begin{table*}[t]
  \centering
  \caption{Open-circle CAPTCHA benchmark ($n=10$, tolerance 30\,px). While most standalone MLLMs correctly identify the target circle, they frequently fail to report sufficiently precise coordinates. The image processing pipeline achieves perfect accuracy with low latency, and hybrid MCP-based configurations inherit this performance, raising all evaluated models to 100\% accuracy.}
  \label{tab:circle-mllm}
  \begin{tabular}{llrrrr}
    \hline
    \textbf{Model} & \textbf{Reasoning} & \textbf{Acc.} &
      \textbf{Avg dist} & \textbf{Avg time} & \textbf{Avg tokens} \\
    \hline
     GPT-4o                	& default 	& 20.0\,\%  	& 104.7 px  	& 1.6\,s  	& 534 \\
    Grok-4.20          	& none    	& 10.0\,\%  	&  97.4\,px  	& 1.3\,s 	& 613\\
    Grok-4.20          	&  default 	& 10.0\,\% 	&  97.4\,px  	& 12.9\,s 	& 2\,321 \\
    Gemini 3.1 Pro 	& low     	& 100.0\,\% 	& 4.0\,px  		& 43.9\,s 	& 1\,308    \\
    Gemini 3.1 Pro 	& default 	& 100.0\,\% 	& 2.1\,px  		& 88.1\,s 	& 2\,607 \\
    \hdashline
    Image processing pipeline & --- & 100.0\,\% &   2.0\,px   & 1.4\,s& ---  \\
     \hdashline
      Hybrid GPT-4o              	& default 	& 100.0\,\%  	&  2.0\,px   	& 5.5\,s  	& 1\,573\\
      Hybrid Grok-4.20          	& none    	& 100.0\,\%   	&  2.0\,px   	& 5.6\,s 	& 995  \\
      Hybrid Grok-4.20          	&  default 	& 100.0\,\%  	&  2.0\,px   	& 24.7\,s	& 1\,956 \\
      Hybrid Gemini 3.1 Pro 	& low     	& 100.0\,\% 	&  2.0\,px   	& 8.6\,s 	& 3\,441 \\
      Hybrid Gemini 3.1 Pro 	& default 	& 100.0\,\% 	&  2.0\,px   	& 18.2\,s 	& 4\,189 \\
    \hline
  \end{tabular}
\end{table*}

%%%% ROTATION CAPTCHA ==========================================================================================================
\subsection{Rotation CAPTCHA}
\label{sec:rotation}

Previous work has reported rotation-based CAPTCHAs as particularly challenging for MLLMs~\cite{wang2025cognitionevaluationdefensemultimodal}. This observation initially surprised us, as the task appears closely related to object recognition and visual reasoning, domains in which modern MLLMs typically perform well.

As a first experiment, we attempted to solve the rotation CAPTCHAs directly using carefully engineered prompts. The models were asked to predict the required clockwise rotation angle in multiples of $45^\circ$. The results are summarized in Table~\ref{tab:rotation-direct}.

\begin{table}
  \centering
  \scriptsize
\caption{Rotation CAPTCHA --- direct angle estimation ($n=10$, tolerance $\pm20^\circ$). The image processing pipeline achieves perfect accuracy while outperforming all standalone MLLMs by several orders of magnitude in runtime. Hybrid configurations (H) inherit the accuracy of the image processing approach for GPT-4o and Grok, whereas Gemini~3.1~Pro occasionally overrides correct algorithmic solutions, resulting in a reduced accuracy of 80\%.}
  \label{tab:rotation-direct}
  \begin{tabular}{llrrr}
    \hline
    \textbf{Model} & \textbf{Acc.} & \textbf{Time} & \textbf{Tokens} \\
    \hline
    GPT-4o          			& 10.0\,\%           & 3.3\,s    & 485    \\
    Grok 4.20-non-reasoning   	& 10.0\,\%           & 2.5\,s    & 470    \\
    Grok 4.20-reasoning  		& 10.0\,\%           & 65.9\,s  & 5\,028 \\
    Gemini 3.1 Pro low     		& 50.0\,\%           & 13.3\,s  & 2\,564 \\
    Gemini 3.1 Pro default 		& 70.0\,\%  	  & 67.0\,s   & 9\,429 \\
     \hdashline
     Image processing pipeline 	& 100.0\,\%   & 0.006\,s& ---  \\
      \hdashline
       H GPT-4o                            	& 100.0\,\%   & 3.1\,s  	& 1\,485   \\
       H Grok-4.20-non-reasoning 	& 100.0\,\%   &  3.5\,s 	& 791  \\
       H Grok-4.20-reasoning   		& 100.0\,\%   & 16.5\,s  	& 1\,329 \\
       H Gemini 3.1 Pro  low      		& 80.0\,\%     & 48.9\,s 	& 10\,416   \\
       H Gemini 3.1 Pro default  		& 80.0\,\%     & 75.1\,s 	&  14\,390\\
    \hline
  \end{tabular}
\end{table}

The results confirm the findings reported in the literature. With the exception of Gemini~3.1~Pro, all evaluated models performed close to chance level. Since eight possible rotations exist, random guessing yields an expected success rate of $12.5\%$. The observed accuracy of approximately $10\%$ therefore indicates that the models derive little practical benefit from their visual understanding when required to estimate the rotation angle directly.

Inspection of the generated reasoning traces provides additional insight. In many cases, the models correctly identified the image fragments and recognized which visual structures should align. This suggests that the primary difficulty lies not in recognizing the relevant objects, but rather in mentally rotating and aligning them. In other words, geometric transformation appears to be a substantially harder task for MLLMs than object recognition.

To test this hypothesis, we reformulated the problem. Instead of requiring the model to predict a rotation angle, we generated eight candidate images corresponding to rotations in $45^\circ$ increments and asked the model to select the correctly aligned image. The results are shown in Table~\ref{tab:rotation-selection}.

\begin{table}
\centering 
\small
\caption{Rotation CAPTCHA solved as a candidate-selection task. Instead of predicting a rotation angle, the models were presented with eight pre-rotated images at $45^\circ$ intervals and asked to select the correctly aligned candidate ($n=10$). Compared to direct angle estimation (Table~\ref{tab:rotation-direct}), all models achieved substantially higher accuracy, supporting the hypothesis that MLLMs can recognize correct alignments more reliably than they can infer the underlying geometric transformation.}
\label{tab:rotation-selection} 
\begin{tabular}{llrrr} \hline \textbf{Model} & \textbf{Acc.} & \textbf{Time} & \textbf{Tokens} \\ 
\hline 
GPT-4o 				& 50.0\,\% & 46.9\,s 		& 5\,910 \\ 
Grok 4.20-non-reasoning 	& 70.0\,\% & 3.7\,s 		& 1\,170 \\ 
Grok 4.20-reasoning         & 60.0\,\% & 71.2\,s 		& 5\,378 \\ 
Gemini 3.1 Pro low 		& 80.0\,\% & 70.5\,s 		& 9\,195 \\ 
Gemini 3.1 Pro default 	& 80.0\,\% & 106.7\,s 	& 12\,152 \\ 
\hline 
\end{tabular} 
\end{table}

The reformulated task yielded substantially higher accuracy across all evaluated models. Interestingly, explicit reasoning provided little to no benefit in this setting. Both Gemini~3.1~Pro configurations achieved identical accuracy, while the Grok~4.20 non-reasoning model even outperformed its reasoning-enabled counterpart, despite requiring only a small fraction of the inference time. These findings suggest that once the geometric transformation problem is reduced to a visual selection task, the challenge becomes primarily one of image comparison rather than multi-step reasoning. The observed improvement therefore supports the hypothesis that MLLMs are generally capable of recognizing the correct alignment when it is presented explicitly, but struggle to infer the underlying geometric transformation themselves. Performance remained below perfect accuracy, however.

\begin{figure}
\centering
\begin{subfigure}[t]{0.22\textwidth}
  \centering
  \includegraphics[width=\linewidth]{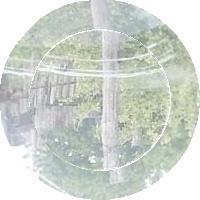}
  \caption{Correct solution. The alignment is slightly imperfect.}
  \label{fig:rotation_off_4}
\end{subfigure}\hfill
\begin{subfigure}[t]{0.22\textwidth}
  \centering
  \includegraphics[width=\linewidth]{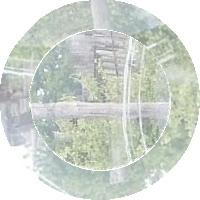}
  \caption{Image selected by Grok-4.20.}
  \label{fig:rotation_off_6}
\end{subfigure}

\caption{Example of an ambiguous rotation CAPTCHA. Grok-4.20 selected candidate 6 instead of the ground-truth candidate 4.}
\label{fig:rotation_off}
\end{figure}

One factor contributing to these residual errors is the nature of the dataset. Since the CAPTCHA samples were collected from real-world deployments, the correct alignment was occasionally imperfect. Furthermore, the image extraction process introduced minor artifacts along the boundary between the inner and outer image regions, as illustrated in Figure~\ref{fig:rotation_off}. In such cases, multiple candidate images may appear visually plausible.

The rotation angle can also be determined reliably using classical image registration techniques. We first extract the pixel values along the boundaries of the inner and outer rings and transform them into one-dimensional angular profiles. To improve robustness against illumination and color variations, we compute radial image gradients rather than operating directly on the raw pixel intensities. The optimal rotational alignment is then estimated by calculating the circular cross-correlation of the resulting gradient signals using the Fast Fourier Transform (FFT). The rotation angle is obtained from the location of the maximum correlation peak, corresponding to the angular offset that best aligns the inner and outer image structures.

This approach proved both substantially faster and more reliable than all evaluated MLLMs. As shown in Table~\ref{tab:rotation-direct}, the image processing pipeline achieved perfect accuracy while requiring only 0.006\,s per challenge. In comparison, the best standalone MLLM reached an accuracy of 70\% while requiring several orders of magnitude more computation time. The performance gap, spanning more than four orders of magnitude, demonstrates that MLLMs are not an efficient solution for geometrically well-defined CAPTCHA challenges. For such tasks, specialized image processing algorithms remain superior in terms of accuracy, runtime, and resource consumption.

This observation mirrors the findings for the open-circle CAPTCHA. In both cases, the underlying challenge consists primarily of a geometric computation rather than a semantic recognition task. Consequently, deterministic image processing methods outperform general-purpose multimodal language models.

The hybrid results reported in Table~\ref{tab:rotation-direct} reveal an important difference compared to the open-circle CAPTCHA. While the GPT-4o and Grok models consistently accepted the rotation angles proposed by the image processing pipeline, Gemini~3.1~Pro occasionally overruled correct solutions. As a result, the hybrid GPT-4o and Grok configurations inherited the perfect accuracy of the image processing pipeline, whereas the hybrid Gemini configurations achieved only 80\% accuracy.

\begin{figure}
\centering
\begin{subfigure}[t]{0.22\textwidth}
  \centering
  \includegraphics[width=\linewidth]{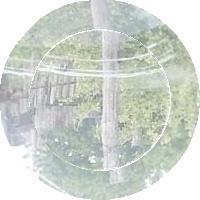}
  \caption{Correct solution, proposed by the image processing pipeline.}
  \label{fig:rotation_overrule_135}
\end{subfigure}\hfill
\begin{subfigure}[t]{0.22\textwidth}
  \centering
  \includegraphics[width=\linewidth]{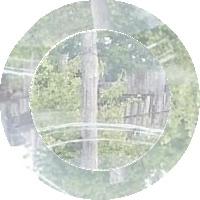}
  \caption{Proposed solution from Gemini 3.1 Pro.}
  \label{fig:rotation_overrule_315}
\end{subfigure}

\caption{Example of a Gemini~3.1~Pro override in the hybrid rotation CAPTCHA solver. The image processing pipeline correctly proposes the discretized CAPTCHA solution of $135^\circ$ (left). However, due to a slight misalignment in the original CAPTCHA, Gemini~3.1~Pro favors an alternative rotation of $315^\circ$ (right) and rejects the proposed solution. This example illustrates how MLLM-based verification can occasionally degrade the performance of an otherwise correct deterministic algorithm when the input contains alignment artifacts.}
\label{fig:rotation_overrule}
\end{figure}

Figure~\ref{fig:rotation_overrule} illustrates a representative example. The image processing pipeline proposed a rotation of 135°, corresponding to the closest valid CAPTCHA solution. However, due to imperfections in the original CAPTCHA generation process, the true optimal alignment was approximately 129°. Gemini~3.1~Pro recognized this slight misalignment and consequently rejected the proposed solution in favor of an alternative rotation. To verify this hypothesis, we repeated the experiment using the exact angle estimated by the image processing pipeline (129° rather than the discretized CAPTCHA solution of 135°). In this setting, Gemini~3.1~Pro accepted all proposed alignments, confirming that the observed failures were caused by the model's sensitivity to minor alignment artifacts rather than by an inability to recognize the correct visual structures.

These findings suggest that MLLM-based verification does not necessarily improve the performance of a reliable deterministic algorithm. For well-defined geometric tasks such as rotation estimation, the image processing pipeline already produces optimal solutions. Introducing an MLLM into the decision loop may increase latency, token consumption, and, in some cases, even reduce accuracy by overriding correct results. Consequently, direct algorithmic solutions remain preferable for narrowly scoped geometric problems, while hybrid architectures are most beneficial when semantic interpretation and tool orchestration are required.

%%%% IMAGE RECOGNITION CAPTCHA  ================================================================================================
\subsection{Image Recognition CAPTCHA}
\label{sec:recognition}

Unlike the open-circle and rotation CAPTCHAs, object-selection CAPTCHAs primarily require semantic image understanding rather than geometric reasoning. Since modern MLLMs are explicitly trained for visual recognition tasks, we expected strong performance on this CAPTCHA category.

A distinctive characteristic of the evaluated insect-selection CAPTCHAs is the heterogeneous nature of the image material. The target insects may appear in photographs, pixel-art representations, realistic illustrations, or highly stylized drawings. Consequently, successful solving requires robust object recognition across multiple visual domains rather than simple pattern matching.

The results are summarized in Table~\ref{tab:insect-results}.

Overall, all evaluated models demonstrated strong insect recognition capabilities. The primary source of errors was not object identification itself, but rather the translation of detections into the requested grid coordinates. The user prompt specified the numbering scheme as follows:

\begin{quote}
\emph{
The grid is numbered from left to right, top to bottom, starting at 1.
}
\end{quote}

The Grok-4.20 non-reasoning model illustrates this failure mode particularly well. In the example shown in Figure~\ref{fig:insects_selection}, the model correctly identified both insect-containing cells but returned the indices \texttt{[5,9]} instead of the correct answer \texttt{[4,9]}. Examination of the accompanying explanation revealed that the model correctly localized the insect spatially, describing it as being in the \emph{middle row, left}, but still assigned an incorrect numerical label. Similar indexing errors were observed throughout the evaluation. In several cases, the model even returned cell numbers outside the valid range (e.g., \texttt{10} in a $3\times3$ grid).

Interestingly, these indexing errors disappeared almost entirely when explicit reasoning was enabled. The Grok-4.20 reasoning model correctly mapped all detected insects to their corresponding grid positions. Its only classification error occurred on a heavily distorted image of a dragonfly, which the model described as a \emph{snake- or worm-like creature}. Given the severe visual distortion, this misclassification is understandable.

A similar trend was observed for GPT-4o. The model correctly numbered all identified objects but failed to recognize two highly distorted insect images among the 90 evaluated cells. In contrast, both Gemini~3.1~Pro configurations achieved perfect performance on the dataset, successfully identifying and correctly indexing all insect-containing cells.

These results suggest that object recognition itself is largely a solved problem for contemporary MLLMs in this setting. The dominant source of failure is instead the conversion of visual observations into a structured response format. Explicit reasoning appears to improve this intermediate reasoning step substantially, resulting in markedly higher overall CAPTCHA-solving accuracy.

\begin{table}
  \small
  \centering
  \caption{Insect-selection CAPTCHA benchmark ($n=10$, 90 image cells). All evaluated models demonstrated strong object recognition capabilities, correctly identifying insects across a diverse set of visual representations. The primary source of errors was not image recognition itself, but the conversion of detections into the required grid indices. Reasoning substantially improved this mapping process, increasing Grok-4.20 accuracy from 10\% to 90\%. Gemini~3.1~Pro achieved perfect performance, indicating that semantic object recognition is largely a solved problem for contemporary MLLMs in this setting.}
  \label{tab:insect-results}
  \begin{tabular}{llrrr}
    \hline
    \textbf{Model} & \textbf{Acc.} & \textbf{Time} & \textbf{Tokens} \\
    \hline
     GPT-4o                     			& 80.0\,\% 	& 2.2\,s  	& 503    \\
     Grok-4.20-non-reasoning          	& 10.0\,\%   	& 8.0\,s   	& 536  \\
     Grok-4.20-reasoning        		& 90.0\,\%  	& 8.5\,s  	& 1\,554  \\
     Gemini 3.1 Pro - low     		& 100.0\,\% 	& 6.2\,s  	& 1\,426 \\
     Gemini 3.1 Pro  				& 100.0\,\% 	& 49.4\,s 	& 1\,560 \\ 
    \hline
  \end{tabular}
\end{table}

\subsection{Summary}

The results obtained for the object-selection CAPTCHA indicate that visual object recognition is largely a solved problem for contemporary MLLMs. All evaluated models were generally able to identify the target insects, even when presented in highly heterogeneous visual styles ranging from photographs to stylized illustrations. The primary source of errors was not object recognition itself, but the conversion of visual observations into a structured response format. In particular, non-reasoning models frequently failed to correctly map detected objects to the required grid indices, whereas reasoning-enabled models performed substantially better.

The results for the rotation CAPTCHA reveal a different limitation. As discussed in Section~\ref{sec:rotation}, even state-of-the-art MLLMs struggled to infer the correct rotational transformation required to align the inner and outer image components. Analysis of the generated reasoning traces suggests that the models were often able to identify the relevant image fragments and determine which structures should align. However, they frequently failed when required to mentally simulate the geometric transformation itself. This finding indicates that the challenge lies not in visual recognition, but in spatial and geometric reasoning.

A similar phenomenon was observed for the open-circle CAPTCHA (Section~\ref{sec:open-circle}). Most models correctly identified the target open circle and could often describe its location qualitatively. Yet they frequently failed to report coordinates with sufficient precision to successfully solve the CAPTCHA. Once again, the limitation was not perceptual recognition, but accurate geometric localization.

Classical image processing techniques provide an effective remedy for these deficiencies. Our experiments demonstrated two complementary strategies:

\begin{itemize}
\item Reformulating the task into a form that is easier for MLLMs to solve. For example, replacing angle estimation with a selection task eliminates the need for mental rotation and substantially improves performance.

\item Augmenting MLLMs with specialized image processing tools capable of performing geometric computations directly. Through MCP integration, the models can delegate tasks such as localization and image registration to deterministic algorithms that solve these problems more accurately and efficiently.
\end{itemize}

Taken together, the results suggest that the strengths of MLLMs and classical image processing are highly complementary. MLLMs excel at interpreting tasks, recognizing semantic content, selecting appropriate tools, and orchestrating complex workflows. In contrast, classical image processing algorithms remain significantly more efficient and reliable for precise geometric computations. A hybrid architecture combines the advantages of both approaches by allowing the MLLM to act as a high-level reasoning and coordination layer while delegating specialized geometric subtasks to dedicated algorithms.

For all CAPTCHA types considered in this study, the hybrid approach enabled reliable and efficient automated solving, demonstrating the practical value of augmenting general-purpose MLLMs with specialized computational tools.

%-------------------------------------------------------------------------------
\bibliographystyle{plain}
\bibliography{\jobname}

\begin{thebibliography}{10}

\bibitem{bradski2008learning}
Gary Bradski and Adrian Kaehler.
\newblock {\em Learning OpenCV: Computer vision with the OpenCV library}.
\newblock " O'Reilly Media, Inc.", 2008.

\bibitem{davies2005machine}
ER~Davies.
\newblock Machine vision: Theory, algorithms, practicalities, 3rd, 2005.

\bibitem{deng2024oedipusllmenchancedreasoningcaptcha}
Gelei Deng, Haoran Ou, Yi~Liu, Jie Zhang, Tianwei Zhang, and Yang Liu.
\newblock Oedipus: Llm-enchanced reasoning captcha solver, 2024.

\bibitem{DengGelei2025OLRC}
Gelei Deng, Haoran Ou, Yi~Liu, Jie Zhang, Tianwei Zhang, and Yang Liu.
\newblock Oedipus: Llm-enchanced reasoning captcha solver.
\newblock In {\em Proceedings of the 2025 ACM SIGSAC Conference on Computer and
  Communications Security}, pages 6--20, New York, NY, USA, 2025. ACM.

\bibitem{duda1973pattern}
Richard~O Duda and Peter~E Hart.
\newblock {\em Pattern Classification and Scene Analysis: Theory and Practice}.
\newblock Wiley, 1973.

\bibitem{GossweilerRich2009WuCa}
Rich Gossweiler, Maryam Kamvar, and Shumeet Baluja.
\newblock What's up captcha?: a captcha based on image orientation.
\newblock In {\em Proceedings of the 18th international conference on World
  wide web}, pages 841--850, New York, NY, USA, 2009. ACM.

\bibitem{GuerarMeriem2022GC’A}
Meriem Guerar, Luca Verderame, Mauro Migliardi, Francesco Palmieri, and Alessio
  Merlo.
\newblock Gotta captcha ’em all: A survey of 20 years of the
  human-or-computer dilemma.
\newblock {\em ACM computing surveys}, 54(9):1--33, 2022.

\bibitem{KumarMohinder2022ASSo}
Mohinder Kumar, M.~K. Jindal, and Munish Kumar.
\newblock A systematic survey on captcha recognition: Types, creation and
  breaking techniques.
\newblock {\em Archives of computational methods in engineering},
  29(2):1107--1136, 2022.

\bibitem{breaking_reCAPTCHA}
Andreas Plesner, Tobias Vontobel, and Roger Wattenhofer.
\newblock Breaking recaptchav2, 09 2024.

\bibitem{309468}
Xiwen Teoh, Yun Lin, Siqi Li, Ruofan Liu, Avi Sollomoni, Yaniv Harel, and
  Jin~Song Dong.
\newblock Are {CAPTCHAs} still bot-hard? generalized visual {CAPTCHA} solving
  with agentic vision language model.
\newblock In {\em 34th USENIX Security Symposium (USENIX Security 25)}, pages
  3747--3766, Seattle, WA, August 2025. USENIX Association.

\bibitem{WangHong2023BoHD}
Hong Wang, Xuan Luo, Weizhi Wang, and Xifeng Yan.
\newblock Bot or human? detecting chatgpt imposters with a single question.
\newblock 2023.

\bibitem{wang2025cognitionevaluationdefensemultimodal}
Junyu Wang, Changjia Zhu, Yuanbo Zhou, Lingyao Li, Xu~He, and Junjie Xiong.
\newblock Cognition: From evaluation to defense against multimodal llm captcha
  solvers, 2025.

\end{thebibliography}

\end{document}